# Observation of two-dimensional lattice interface solitons

A. Szameit[1], Y. V. Kartashov[2], F. Dreisow[1], M. Heinrich[1], V. A. Vysloukh[2], T. Pertsch[1], S. Nolte[1], A. Tünnermann[1,3], F. Lederer[4], and L. Torner[2]

[1]*Institute of Applied Physics, Friedrich-Schiller-University Jena, Max-Wien-Platz 1, 07743 Jena, Germany*

[2]*ICFO-Institut de Ciencies Fotoniques, and Universitat Politecnica de Catalunya, Mediterranean Technology Park, 08860 Castelldefels (Barcelona), Spain*

[3]*Fraunhofer Institute for Applied Optics and Precision Engineering, Albert-Einstein-Strasse 7, 07745 Jena, Germany*

[4]*Institute for Condensed Matter Theory and Optics, Friedrich-Schiller-University Jena, Max-Wien-Platz 1, 07743 Jena, Germany*

We report on the experimental observation of two-dimensional solitons at the interface between square and hexagonal waveguide arrays. In addition to the different symmetry of the lattices, the influence of a varying refractive index modulation depth is investigated. Such variation strongly affects the properties of surface solitons residing at different sides of the interface.

High-intensity excitations propagating along an interface between different nonlinear materials emerge in various areas of optics. It was demonstrated that optical surface waves can be excited at moderate power levels at the interface between a uniform material and a waveguide array with focusing nonlinearity [1,2], at the interfaces between dissimilar arrays [3,4], or at the edge of defocusing lattices [5-7]. Quadratic surface solitons were observed also [8]. Surface soliton families become richer at two-dimensional (2D) geometries [9,10]. They were observed in optically induced lattices [11] and in laser-written waveguide arrays [12]. To date, in 2D settings experiments were conducted for interfaces between uniform and periodic media. However, the interface of different periodic materials may support more exotic states [3,4,13,14]. In this Letter we report on the first experimental observation of 2D surface solitons at interfaces of arrays with different symmetry (hexagonal and square) and with different refractive index modulation depths. Such surface solitons feature asymmetric shapes, while differences in array properties affect strongly the threshold power for soliton existence and excitation.

To gain physical insight into the properties of surface solitons in different arrays we conducted a theoretical analysis assuming cw illumination. We describe the beam propagation by the Schrödinger equation for the dimensionless field amplitude $q$:

$$i\frac{\partial q}{\partial \xi} = -\frac{1}{2}\left(\frac{\partial^2 q}{\partial \eta^2} + \frac{\partial^2 q}{\partial \zeta^2}\right) - q|q|^2 - pR(\eta,\zeta)q. \qquad (1)$$

Here $\eta,\zeta$ are the transverse and $\xi$ is the longitudinal coordinate; the refractive index modulation depth is determined by $p = p_s$ in the square array and $p = p_h$ in the hexagonal one; $R(\eta,\zeta)$ is the refractive index profile given by a superposition of Gaussian functions $\exp[-(\eta/w_\eta)^2 - (\zeta/w_\zeta)^2]$ describing profiles of individual guides with spacing $d$. In our samples with focusing nonlinearity ($n_2 = 2.7 \times 10^{-20}$ m$^2$/W) the guides exhibit transverse dimensions $4.5 \times 9$ $\mu$m$^2$ and the waveguide spacing is $40$ $\mu$m. Thus, we set $w_\eta = 0.45$, $w_\zeta = 0.9$, and $d = 4$ using the transverse scale $x_0 = 10$ $\mu$m; $p = 1$ is equivalent to a refractive index change of $1.1 \times 10^{-4}$. To observe the influence of the interface, we consider excitations residing in the first rows of square or hexagonal arrays [Fig. 1(c)].

We search for soliton solutions in the form $q = w(\eta,\zeta)\exp(ib\xi)$, characterized by the power $U = \int\int_{-\infty}^{\infty} w^2 d\eta d\zeta$. Surface solitons exist above a power threshold $U_{\text{th}}$ and above a cutoff $b_{\text{co}}$ on propagation constant $b$ (Fig. 1). When waveguides in both arrays have equal refractive index ($p_s = p_h = 3$), the power thresholds for solitons residing at the marked positions are close and amount to $U_{\text{th}}^{\text{s}} = 0.652$ in the square part and $U_{\text{th}}^{\text{h}} = 0.685$ in the hexagonal part [Fig. 1(a)]. The soliton penetrates deeper into the hexagonal array due to higher packing density of waveguides in this array leading to higher mean refractive index. This shape asymmetry is most pronounced close to the cutoff and is further enhanced when the refractive index in the hexagonal array is higher ($p_s = 3$, $p_h = 3.2$). In this case solitons in square array which are close to the cutoff acquire characteristic three-spot shape [Fig. 2(a)], while far from the cutoff they are localized [Fig. 2(b)]. In contrast, solitons residing in the hexagonal array strongly expand into this array, but do not penetrate into square array [Fig. 2(c)]. This picture changes completely when the refractive index is higher in square array ($p_s = 3.2$, $p_h = 3.0$). In this case solitons residing in hexagonal array tend to penetrate into square array [Fig. 2(d)], while weakly localized solitons in square array strongly expand into this region and do not penetrate into hexagonal array [Fig. 2(f)].

The refractive index change in one of the arrays profoundly affects the threshold power. While at $p_s = p_h = 3$ the thresholds for solitons in the square and hexagonal arrays are comparable ($U_{th}^s = 0.652$ and $U_{th}^h = 0.685$), at $p_h = 3.2 > p_s = 3.0$ the threshold for solitons in hexagonal array *decreases* to $U_{th} = 0.537$, while the threshold for soliton formation in square array *increases* to $U_{th} = 0.887$ [Fig. 1(a)]. When $p_s = 3.2 > p_h = 3.0$ the threshold power in square array *decreases* to $U_{th}^s = 0.441$, while it *increases* to $U_{th}^h = 0.873$ in hexagonal array [Fig. 1(b)]. Thus, the excitation of surface solitons inside the shallower array requires substantially *higher* powers than the excitation of the same soliton at the interface of arrays of equal depths. In contrast, it becomes *easier* to excite solitons inside deeper array. This is because when $p_h > p_s$, a beam launched in the square array tends to penetrate into the hexagonal array with higher refractive index, while a surface wave can only form when nonlinearity balances the refractive indices at both sides of the interface. Thus, higher threshold powers are necessary to prevent tunneling into the hexagonal array with higher $p_h$. In contrast, upon excitation in the hexagonal array, light weakly penetrates into the square array and the threshold power is determined by the refractive index of hexagonal array and decreases with increase of $p_h$. This regime holds when the refractive index difference between the arrays is sufficiently small.

To confirm the theoretical predictions, we fabricated arrays consisting of hexagonal and square regions with a length of 105 mm (see [15] for details of fabrication). The writing velocity was 2000 $\mu$m/s, which ensures a nonlinear coefficient similar to that in the bulk material [16]. Due to the large spacing (40 $\mu$m) the evanescent coupling is almost isotropic. The transmission losses of a single waveguide were $< 0.4$ dB/cm. For soliton excitation we used a Ti:Sa CPA laser system (Spitfire, Spectra-Physics) with a pulse duration of 150 fs and a repetition rate of 1 kHz at 800 nm.

Figure 3 shows the experimental output patterns for our three samples (in comparison with simulations), when a waveguide in the first row in the *square* array is excited with an input peak power of 3.2 MW. In Fig. 3(a) both array parts exhibit the same refractive index modulation depth, in Fig. 3(b) the index of the square array part is increased due to reduced writing speed of 1800 $\mu$m/s, while in Fig. 3(c) the hexagonal region exhibits a larger index. In contrast to Figs. 3(a) and 3(b), where the light is strongly localized, in Fig. 3(c) one observes only an intermediate localization, when light penetrates into the hexagonal region. Note that experiments are performed with light pulses, under conditions where dispersion effects are negligible, while the previous analysis was conducted assuming cw illumination. Therefore, the experimentally observed threshold power for soliton formation exceeds slightly the theoretical estimate obtained for cw radiation, because even when the peak power exceeds the cw soliton threshold, the pulse wings still diffract yielding a broadened time-integrated background.

When exciting a waveguide in the first row of the *hexagonal* array, strong localization is observed if the waveguide indices in the hexagonal and square regions are equal [Fig. 4(a)] and when the refractive index in the hexagonal array is higher [Fig. 4(c)]. In Fig. 4(b) only an intermediate localized state is observed, since light strongly penetrates into the square array due to its higher refractive index. Hence, the effects of increased refractive index in the square array dominate over effects arising due to higher packing density of waveguides in hexagonal array. This behavior is consistent with the numerical simulations (bottom row of Fig. 4).

A particular feature of light propagation at interfaces between two different structures is the enhanced light penetration into the region with higher refractive index. A sequence of output intensity distributions for increasing input peak powers is depicted in Fig. 5, where a waveguide in the first row of the square array was excited and the hexagonal array had a stronger refractive index change than the square one. For low peak powers the light is almost confined to the square region since it is reflected at the interface [Fig. 5(a)]. For increasing power the light starts to penetrate into the hexagonal region [Fig. 5(b) and 5(c)] because nonlinearity increases the refractive index and causes phase matching of both array sections. However, for large enough input peak powers one can clearly observe near-surface localization [Figs. 5(d) and 5(e)], so that finally a surface lattice soliton forms [Fig. 5(f)].

Thus, summarizing, we observed the formation of two-dimensional surface lattice solitons at the interface between square and hexagonal waveguide arrays written by a fs-pulse technique. Our observations show that the penetration of the light into structure significantly depends on lattice symmetry, the difference of the refractive index of the two regions and the applied input power.

# References with titles

# References without titles

## Figure captions

Figure 1. Power $U$ versus $b$ for (a) $p_{\rm s}=3$, $p_{\rm h}=3.2$, and (b) $p_{\rm s}=3.2$, $p_{\rm h}=3$. Curves labeled $U_{\rm s}(U_{\rm h})$ correspond to solitons residing in square (hexagonal) array. The circles in (a) and (b) correspond to solitons shown in Figs. 2(a)-2(c) and 2(d)-2(f), respectively. Red curves in (a) and (b) show $U_{\rm s}(b)$ for $p_{\rm s}=p_{\rm h}=3$. (c) Microscope image of a laser written array with marked excited waveguides and vertical line indicating interface position.

Figure 2. Surface solitons at (a) $b=0.594$, (b) $0.657$, (c) $0.589$, (d) $0.587$, (e) $0.643$, and (f) $0.570$. In (a)-(c) one has $p_{\rm s}=3$, $p_{\rm h}=3.2$; in (d)-(f) $p_{\rm s}=3.2$, $p_{\rm h}=3$.

Figure 3. Comparison of the output intensity distributions for an excitation of a waveguide in the first row of the square array, when (a) $p_{\rm s}=p_{\rm h}$, (b) $p_{\rm s}>p_{\rm h}$, and (c) $p_{\rm s}<p_{\rm h}$. Top row - experiment, bottom row - theory. In all cases the input power is $3.2$ MW.

Figure 4. The same as in Fig. 3 but for an excitation of a waveguide in the first row of the hexagonal array.

Figure 5. Dynamic excitation of a waveguide in the first row of the square array, when $p_{\rm s}<p_{\rm h}$. The input power is (a) $62$ kW, (b) $1.4$ MW, (c) $2$ MW, (d) $2.3$ MW, (e) $2.7$ MW, and (f) $3.2$ MW.

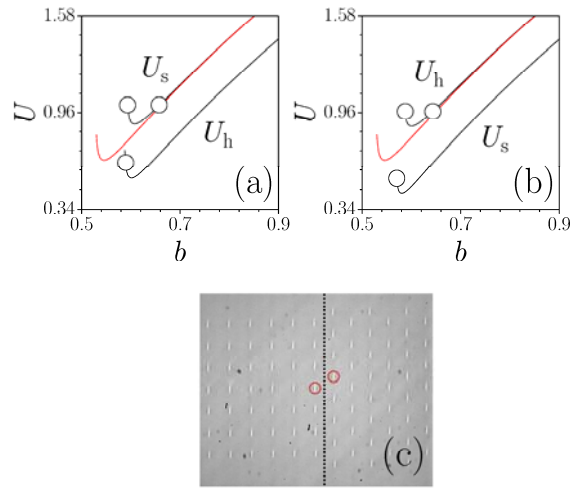

Figure 1. Power $U$ versus $b$ for (a) $p_\text{s}=3$, $p_\text{h}=3.2$, and (b) $p_\text{s}=3.2$, $p_\text{h}=3$. Curves labeled $U_\text{s}(U_\text{h})$ correspond to solitons residing in square (hexagonal) array. The circles in (a) and (b) correspond to solitons shown in Figs. 2(a)-2(c) and 2(d)-2(f), respectively. Red curves in (a) and (b) show $U_\text{s}(b)$ for $p_\text{s}=p_\text{h}=3$. (c) Microscope image of a laser written array with marked excited waveguides and vertical line indicating interface position.

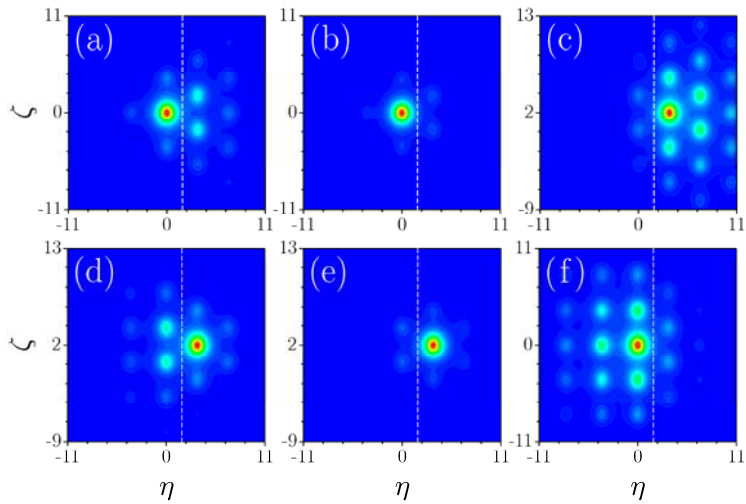

Figure 2. Surface solitons at (a) $b = 0.594$, (b) $0.657$, (c) $0.589$, (d) $0.587$, (e) $0.643$, and (f) $0.570$. In (a)-(c) one has $p_s = 3$, $p_h = 3.2$; in (d)-(f) $p_s = 3.2$, $p_h = 3$.

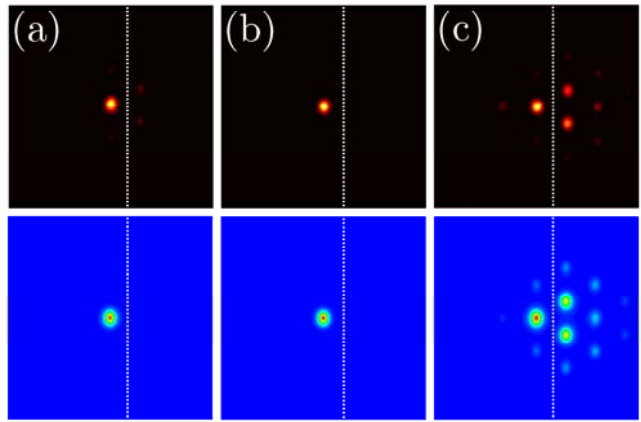

Figure 3. Comparison of the output intensity distributions for an excitation of a waveguide in the first row of the square array, when (a) $p_\text{s} = p_\text{h}$, (b) $p_\text{s} > p_\text{h}$, and (c) $p_\text{s} < p_\text{h}$. Top row - experiment, bottom row - theory. In all cases the input power is 3.2 MW.

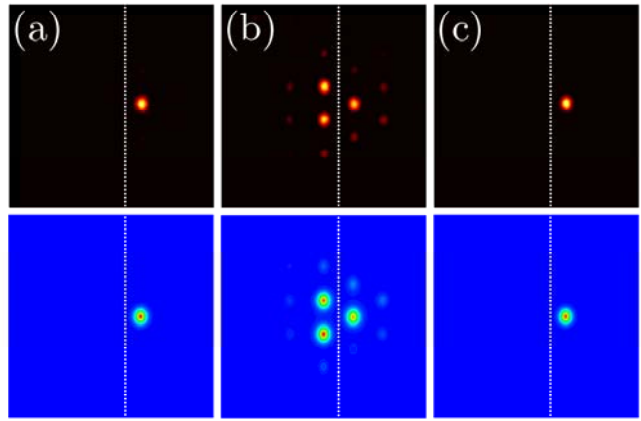

Figure 4. The same as in Fig. 3 but for an excitation of a waveguide in the first row of the hexagonal array.

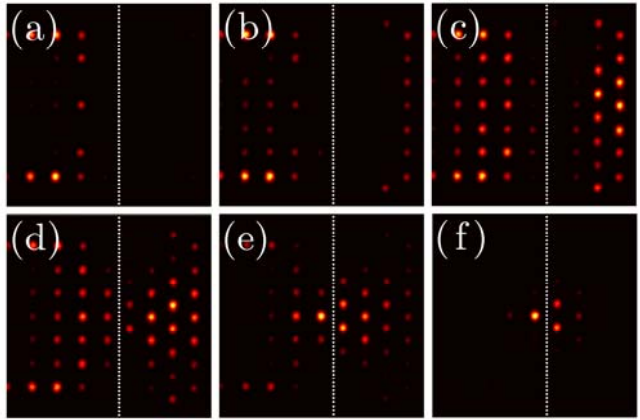

Figure 5. Dynamic excitation of a waveguide in the first row of the square array, when $p_s < p_h$. The input power is (a) 62 kW, (b) 1.4 MW, (c) 2 MW, (d) 2.3 MW, (e) 2.7 MW, and (f) 3.2 MW.